\documentclass[10pt,letterpaper]{article}
\usepackage{opex3}
      
\newcommand{\bra}[1]{\left\langle #1\right|}
\newcommand{\ket}[1]{\left| #1\right\rangle}

\usepackage{amssymb,amsmath,amsthm,amsfonts}
\usepackage{graphicx}
\usepackage{epsfig}
\usepackage{subfigure}
\usepackage{amsmath}
\usepackage{dsfont}

\begin{document}

\title{Long-distance practical quantum key 
  distribution by entanglement swapping}

\author{Artur Scherer, Barry C.\ Sanders, Wolfgang Tittel}

\address{Institute for Quantum Information Science, University of Calgary, \\ Calgary, AB, Canada T2N 1N4}

\email{ascherer@ucalgary.ca} 



\begin{abstract}
We develop a model for practical, entanglement-based  
long-distance quantum key distribution 
employing entanglement swapping as a key building block.
Relying only on existing off-the-shelf~technology, 
we show how to optimize resources so as to maximize     
secret key distribution rates. The tools comprise lossy transmission 
links, such as telecom optical fibers or free space,
parametric down-conversion sources of entangled photon pairs,   
and threshold detectors that are inefficient and have dark counts. 
Our analysis provides the optimal trade-off between detector efficiency 
and dark counts, which are usually competing, as well as the optimal source brightness 
that maximizes the secret key rate for specified 
distances (i.e.\ loss) between sender and receiver.
\end{abstract}



\vspace*{1mm}
\section{Introduction}

Quantum cryptography technologies have matured to  
commercial applications~\cite{QKD-Companies}. Yet long-distance quantum communication and 
particularly long-distance quantum key distribution (QKD) 
are hampered by exponential channel loss of photons with respect to transmission distance. 
Quantum relays~\cite{Gisin2002,EdoWaks2002,Jacobs2002,Riedmatten2004,Collins2005} and quantum
repeaters~\cite{BriegelDuerCiracZoller1998} could solve these distance limits by exploiting  
entanglement swapping (ES)~\cite{Zukowski1993} between photon pairs 
as a key building block. However, ES based on present technology 
is performance-limited due to real-world imperfections.

ES is achieved with two sources of entangled photon pairs (EPPs) and a joint  
Bell-state measurement (BSM) performed on two of their outputs, specifically one from
each source. Realistic  EPP sources are probabilistic,  
and occasionally emit two or more independent EPPs.  
Spontaneous parametric down-conversion (PDC) in nonlinear 
crystals is the  most common 
way to produce EPPs. Detectors used to
perform BSM are inefficient and suffer from dark counts. 
Aiming at realistic aspects of ES-based QKD, 
we have studied the effect of experimental imperfections on   
ES-generated entangled quantum states (in terms of their fidelity 
with a target Bell state) via a {\em non-perturbative}    
mathematical model for practical ES accounting for  
detector inefficiencies, detector dark counts and 
multipair events~\cite{Scherer2009}. 
Our closed-form solution for realistic ES-generated quantum states determines 
the \lq\lq amount''  of useful entanglement, which depends on experimental
parameters such as dark-count rates
and efficiencies of off-the-shelf detectors as well as brightness of
PDC sources. This realism makes our model useful for planning long-distance QKD
experiments employing ES, which is demonstrated in this paper.  

The impact of real-world imperfections on the performance and 
communication range of QKD has been the objective of numerous recent
investigations~\cite{Scarani2009}.
In~\cite{Brassard2000} Brassard et al.\ showed that channel losses, 
a realistic detection process, and qubit-source imperfections 
drastically impair the feasibility of QKD over long distances. 
In particular, it was shown in~\cite{Brassard2000} that unconditional security is difficult 
to achieve in long-distance QKD based on the BB84 protocol~\cite{BB84} 
realized by attenuated laser pulses instead 
of by idealized single-photon on-demand sources; in the same work, a superior 
performance was obtained for QKD schemes based on a single PDC source.  
The consequences of using probabilistic EPP sources (realized by
PDC) instead of by single-pair on-demand sources for quantum 
communication including entanglement-based QKD have been investigated~\cite{Ma2007,Marcikic2002,Riedmatten2003}. 

For long distances, the entanglement-based BBM92 QKD protocol~\cite{BBM92} with a
single PDC source placed midway between the two communicating parties was shown to
perform significantly better than BB84-based QKD realized 
by faint coherent-state pulses under the restriction to individual eavesdropping
attacks and trusted noisy detectors~\cite{EdoWaks2002}. 
Even though faint-pulse BB84 QKD with {\em decoy states} 
(decoy-BB84 QKD)~\cite{Hwang2003,WangPRL2004,Lo2005-1,Lo2005-2,WangPRA2005,Lo2006} 
permits much larger communication ranges than conventional faint-pulse BB84 QKD
without decoy states, PDC-based BBM92 QKD with the source in the middle 
tolerates higher channel loss and thus enables longer communication 
distance than decoy-BB84 QKD~\cite{Ma2007},  setting aside the fact that 
the latter protocol can realize appreciably higher key-distribution rates than the 
former for medium- and low-loss settings.
Moreover, for {\em ideal} EPP sources, 
ES-based BBM92 QKD schemes were proven to allow achieving even greater distances 
at the cost of smaller communication rates~\cite{EdoWaks2002,Collins2005}. 

Here we extend our entanglement-swapping model~\cite{Scherer2009} to practical 
QKD based on distributing entangled photons 
over extended distances by ES and relying 
only on existing off-the-shelf technology. 
The resources we consider are (i)~lossy transmission links, 
such as telecom optical fibers or free space, 
(ii)~spontaneous PDC to produce EPPs  
and (iii)~inefficient, noisy threshold detectors.  
We show how to employ these resources so as to optimize QKD performance.
We determine the QKD figures of merit  {\em quantum bit error rate} 
({\small\sc QBER}) and {\em secret key rate} as functions of 
experimental parameters. 
Our theory permits constrained optimization 
of the experimentally tunable detector efficiencies and
dark count rates as well as brightness of PDC thereby yielding optimal QKD 
performance for any  distance \emph{d} between sender  Alice and  receiver
Bob. 

Determining optimal source brightness is important for both 
faint-pulse BB84 QKD and PDC-based BBM92 QKD. Low source brightness implies 
a low key-generation rate. However, as the source brightness 
increases, the multiphoton-signal probability for 
faint-pulse BB84 QKD or multipair probability for PDC-based BBM92 QKD rises.  
In faint-pulse BB84 QKD, multiphoton signals are vulnerable to    
photon number splitting (PNS) attacks which jeopardize QKD 
security. 
Decoy states~\cite{Hwang2003,WangPRL2004,Lo2005-1,Lo2005-2}) are generally used 
as a remedy to tackle this problem. Although PNS  attacks do not 
help an eavesdropper in PDC-based BBM92 QKD~\cite{Gisin2002,Collins2005}, 
occasional multipairs cause erroneous heralding events, thus contributing 
to {\small\sc QBER} (see
Sec.~\ref{Resources}). 

The optimal mean photon number per signal for faint-pulse coherent-state QKD 
can be determined~\cite{Lo2005-2,Luetkenhaus2000} as well as 
the optimal source brightness for PDC-based BBM92 QKD 
with a single PDC source placed midway between sender and receiver~\cite{Ma2007}.
Although the effects of transmission losses, 
detector inefficiencies and dark
counts on the performance of quantum relays has been examined~\cite{Collins2005}, 
the probabilistic nature of realistic EPP sources including 
occasional multipair events has not yet been 
incorporated. Our analysis yields 
optimal PDC-source brightness for 
PDC-based BBM92 QKD exploiting ES as a crucial tool (PDC-ES-BBM92 QKD) 
for any  channel distance given an empirical constraint between 
efficiency and dark counts for common off-the-shelf detectors. 

Multi-excitation events, as significant sources of
error, have been considered in other recent investigations that elaborate   
on practical implementations of the DLCZ 
quantum repeater scheme~\cite{Duan2001}. 
The effects of multipair events in PDC, in addition to transmission
losses, detector and quantum memory imperfections on quantum repeater
performance, have been accounted for perturbatively~\cite{BraskSorensen2008}. 
The atom-light entangled states produced by Stokes scattering in
the DLCZ scheme are similar to the light-light entangled states produced 
by non-degenerate PDC, as both are two-mode squeezed states. 
In these works~\cite{JiangTaylorLukin2007,Zhao-et-al2007,Brask2010} 
the impact of multi-excitation contributions in 
such atom-light entangled states has also been taken into account perturbatively. 
A more thorough analysis of multi-excitation events in atomic-ensemble
memories has been provided in~\cite{Amirloo2010}. Our theory is based 
on a substantially different approach, which is 
non-perturbative and uses the principle of Bayesian inference 
to account for the presence of experimental imperfections.

Finally, we address the communication range and 
corresponding key generation rates achieved by 
PDC-ES-BBM92 QKD compared to decoy-BB84 QKD. 
In particular, we analyze the conjecture that there is no 
superiority of PDC-ES-BBM92 QKD over decoy-BB84 QKD  
with respect to achievable range for detectors with 
negligible dark counts. 

This paper is organized as follows. In Sec.~\ref{Resources} 
we identify the resources and describe how we incorporate 
real-world imperfections into our mathematical model. 
In Sec.~\ref{Sec:OptimizingQKD} we demonstrate how to 
optimize the performance of PDC-ES-BBM92 QKD 
with respect to PDC sources and detectors. 
Sec.~\ref{Sec:ComparisonPDC-ES-BBM92vsDecoy-BB84QKD} 
provides a comparison between PDC-ES-BBM92 QKD and 
decoy-BB84 QKD for negligibly small detector dark count 
rates. We conclude in Sec.~\ref{Sec:Conclusion} with 
a brief summary and important remarks.

\section{Identifying the resources}\label{Resources}

The photon transmission probability is $10^{-\alpha l/10}$ for  
$\alpha$ the loss coefficient (in dB$/$km) and 
$l$ the distance the light travels. 
The loss differs depending on whether transmission is via 
fiber optics or free space. In this analysis 
we leave the loss coefficient unspecified and normalize the distance 
between sender and receiver through the product $\alpha d$. 
For example, the loss coefficient for light of wavelength $1550$ nm 
propagating through a telecom optical fiber is approximately $\alpha\approx
0.25$ dB km$^{-1}$ \cite{Collins2005}, so $\alpha d=10$ (a \lq\lq $10$ dB
loss'') corresponds to $d\approx 40$ km of fiber. 

\begin{figure}[tb]
\begin{center}
\includegraphics[width=0.97\linewidth]{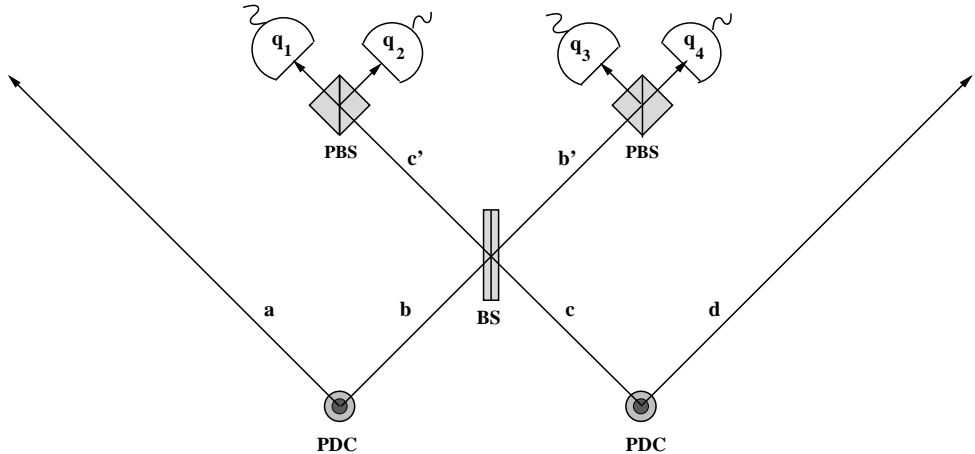}
\end{center}
{\caption{
Entanglement swapping based on two PDC sources and an interferometric BSM. 
Four spatial modes are involved: $a$, $b$, $c$ and $d$. 
The modes $b$ and $c$ are combined at a balanced beam\-splitter (BS). 
Outputs $b'$ and $c'$ are directed to 
polarizing beamsplitters (PBS) and then detected 
at four photon detectors: one for the $H$ and one for the $V$ polarization 
of each of the $c'$ and $b'$ modes. The detectors are inefficient 
and subject to dark counts. Their readout is denoted by $\{q_1,q_2,q_3,q_4\}$. 
In a QKD experiment (see Fig.~\ref{Fig:AliceRelayBob}), 
the polarization-entangled photons of the remaining modes $\,a\:$ and $\,d\;$ 
are distributed between Alice and Bob, respectively, and the BBM92 protocol
\cite{BBM92} can be applied to make the 
secret key. 
\label{Fig:Setup}}}
\end{figure}

Basic experimental ES is illustrated in Fig.~\ref{Fig:Setup}. 
Two PDC sources emit photon pairs into spatial modes  
$a$ and $b$ (first PDC) and $c$ and $d$ (second PDC).
For ES, a joint BSM 
is performed on the  $b$ and $c$ modes. 
As a consequence, provided that a certain 
measurement readout occurs, the photons in  
outgoing  modes $a$ and $d$ emerge entangled 
despite never having interacted with 
one another~\cite{Zukowski1993}. 
The entanglement previously contained in the 
$a$ and $b$ and the $c$ and $d$ photon pairs, respectively, is 
swapped to the $a$ and $d$ photon pair.

We assume type-I nondegenerate PDC as the generator of 
{\em polarization-entangled} quantum states 
\begin{equation}
\ket{\chi}_{ab}=\exp[i\chi(\hat{a}_{\mbox{\tiny{H}}}^{\dagger}
\hat{b}_{\mbox{\tiny{H}}}^{\dagger}+\hat{a}_{\mbox{\tiny{H}}}
\hat{b}_{\mbox{\tiny{H}}})]\otimes
\exp[i\chi(\hat{a}_{\mbox{\tiny{V}}}^{\dagger}
\hat{b}_{\mbox{\tiny{V}}}^{\dagger}+\hat{a}_{\mbox{\tiny{V}}}
\hat{b}_{\mbox{\tiny{V}}})]\ket{\mbox{vac}}
\end{equation}
in two spatial modes $a$ and $b$, 
where $\ket{\mbox{vac}}$ is the multimode va\-cuum state.
Our analysis is straightforward to generalize to other types 
of entanglement. 
The parameter $\chi\in\mathbb{R}$ is proportional to the 
$\chi^{(2)}$ nonlinearity of the crystal, the strength of the 
pump laser and the interaction time between the 
pump laser and the medium, which is approximately   
the laser-pulse duration. 
The value of $\chi^2$ (the square of $\chi$, not to 
be confused with  the crystal's $\chi^{(2)}$ nonlinearity) is the probability for 
EPP generation within the time window 
of a laser pulse. Equivalently, $\chi^2$ can be 
interpreted as the EPP production rate (brightness) of
the PDC source. We assume the same brightness for both PDCs. 
The quantum state generated by the second PDC source 
is thus $\ket{\chi}_{cd}$, and the common quantum state prepared 
by two identical PDC sources is then given by 
$\ket{\chi}_{abcd}=\ket{\chi}_{ab}\otimes\ket{\chi}_{cd}$. 
Exceedingly  small $\chi$ values imply disadvantageously 
low EPP production rates; however, as $\chi$ increases, 
the probability for harmful multipair events rises, which lead  
to faulty detection clicks and thus incorrect estimates of 
entanglement after~ES. 

We model detector efficiency $\eta$ ($0\le\eta\le 1$) 
by preceding a fictitious unit-efficiency, dark-count-exempt photon-counting 
detector with a virtual beamsplitter 
of transmittance $\eta$. We include dark counts       
by a fictitious thermal background source of light 
incident on the second input port of the beamsplitter.  
Dark counts are incorporated by the 
dark count probability $\wp_{\mbox{\tiny{dc}}}$ 
per time window of the pump laser pulse in the PDC process. 
In our model transmission losses are included in detector  
efficiencies according to  $\eta=\eta_0 10^{-\alpha l/10}$, where $\eta_0$
denotes the intrinsic detector efficiency. 

Detector quality is important for long-distance QKD.
On the one hand, detectors should be as efficient as possible 
to achieve fast key rates and high efficiency for ES operation. On the
other hand, dark count noise should be low to make the 
{\small\sc QBER} small.  
In experiments these two aims compete: $\eta_0$ and $\wp_{\mbox{\tiny{dc}}}$ 
counteract for typical off-the-shelf detectors. 
Avalanche photodiodes (APDs) are the most used photon detectors in 
long-distance quantum communication experiments over telecom optical fibres, 
with InGaAs/InP diodes being the most common. The trade-off between efficiency and dark
counts 
for high quality InGaAs APD detectors at optical telecom wavelength 
 $1550$ nm  
can be characterized by the empirical relation 
\begin{equation}
\wp_{\mbox{\tiny{dc}}}=A\exp(B\eta_0)
\label{Eq:constraint}
\end{equation}
with typical values $ A=6.1\times 10^{-7}$ and
$B=17$~\cite{Collins2005}.
For simplicity, we assume the same efficiency $\eta_0$ and dark count probability
$\wp_{\mbox{\tiny{dc}}}$, 
respectively, for all
detectors employed in the here considered QKD scheme, 
subject 
to the empirical constraint (\ref{Eq:constraint}). 

In fact single-photon detectors other than APDs 
have been prototyped. Most notable are   
superconducting transition-edge
sensors (TES), 
which are photon-number-resolving with up to 88\% detection efficiency at  
$1550$ nm and benefit from negligible dark count
rates~\cite{Miller2003,Rosenberg2005A,Rosenberg2005B}. 
A further type with a demonstrated photon-number-resolving functionality 
is given by superconducting nanowire single-photon detectors
(SNSPDs)~\cite{Nanowires-1,Nanowires-2,Nanowires-3,Nanowires-4}, 
which combine a high infrared detection efficiency (up to 57\% 
at $1550$ nm~\cite{Nanowires-2}) with an ultra-low dark count rate 
and a high counting frequency~\cite{Nanowires-4}. Such detectors 
(TES or SNSPDs) substantially increase the range, security and bit rate for QKD. 
However, a severe drawback of both TES and SNSPDs is the fact that they must be operated 
at cryogenic temperatures, which makes them impractical for 
off-the-shelf QKD technology.

Transmission of polarization-encoded qubits naturally suffers from 
depolarization, i.e., environment-induced randomization of 
photon polarization. The amount of this depolarization depends 
on the properties of the quantum channel, i.e., on the specific fiber 
used for photon transmission or atmospheric conditions in 
the case of free-space QKD, as well as on the spectral band-width of 
the individual photons. Nevertheless, modern fibers affect  
the polarization far less than previously thought.
High-fidelity transmission of polarization-encoded 
qubits from EPP sources is possible and was successfully 
demonstrated over 100 km of fiber~\cite{HuebelOptExp2007} 
and in free space even up to 144 km~\cite{Zeilinger-NaturePhysics2007}.
Such high-fidelity transmission can be extended to even greater 
distances by spectral filtering of the down-converted 
photons~\cite{Halder-Nature2007,Saglamyurek2011}. 
Moreover, various implementations have demonstrated how 
to remedy birefringence-caused, time-varying unitary polarization 
transformations during photon transmission. 
Promising proposals include, e.g., real-time polarization 
control employing two nonorthogonal reference signals 
multiplexed in either time or wavelength with the data 
signal~\cite{Xavier2008} as well as stabilization of unwanted 
qubit transformation in the quantum channel using quantum 
frames~\cite{Martinez2009}. Hence, for distances up to 100 km, 
and probably beyond this range, the degree of observed quantum 
correlations is limited mainly by  detector dark counts and 
multi-pair emissions rather than by depolarization, 
which we neglect in the present analysis.

Previously we derived a nonperturbative, closed-form 
solution for the quantum states $
\hat{\rho}^{\{q_\nu\}}\left(\chi,\{\eta_{\nu}\},\{{\wp_{\mbox{\tiny{dc}}}}_{\nu}\}\right)$  
prepared by a realistic ES, given a recorded readout  $\{q_\nu\}$ (e.g.\
$\{q_1,q_2,q_3,q_4\}$ in Fig.~\ref{Fig:Setup})
of a BSM with faulty detectors characterized by efficiencies $\{\eta_\nu\}$ and  
dark count probabilities $\{{\wp_{\mbox{\tiny{dc}}}}_{\nu}\}$ 
($\nu$~is a label for different detectors involved in 
the BSM), as a density-operator valued function of $\chi$, $\{\eta_{\nu}\}$ and 
$\{{\wp_{\mbox{\tiny{dc}}}}_{\nu}\}$~\cite{Scherer2009}. Using this closed-form solution, we 
can simulate a four-fold coincidence experiment.  
A direct measure for entanglement quantification after ES  
is the  visibility 
$V:=(\mbox{\small \sc Max}-{\mbox{\small \sc
    Min}})/(\mbox{\small \sc
  Max}+{\mbox{\small \sc Min}})$,  
 where \lq\lq $\mbox{\small\sc Max}$'' and \lq\lq $\mbox{\small\sc Min}$'' 
denote the maximum and minimum values of 
the four-fold coincidence rate as a function of polarization angle. 
Provided that click events are observed in both the $a$ and $d$ modes, 
and restricting ourselves to the corresponding post-selected quantum states 
$ \hat{\rho}^{\{q_\nu\}}_{\mbox{\tiny
    postsel}}$, the visibility is directly connected to the fidelity  
$F=\bra{\psi^T}\hat{\rho}^{\{q_\nu\}}_{\mbox{\tiny
    postsel}}\ket{\psi^T}$  
with respect to a target Bell state $\ket{\psi^T}$ via the relation 
$V=(4F-1)/3$~\cite{RiedmattenMarcikicHouwelingenTittelZbindenGisin2005}.
The relation between visibility and correlation coefficient $S_{\mbox{\tiny CHSH}}$ of the CHSH Bell
inequality is $S_{\mbox{\tiny CHSH}}=2\sqrt{2}\,V$, cf.~\cite{Rarity1990}.  
Our predictions~\cite{Scherer2009} agree with experimental results~\cite{RiedmattenMarcikicHouwelingenTittelZbindenGisin2005}: 
our theory predicts $V_{\mbox{\tiny theory}}=77.7\%$, and the 
observed visibility in experiment was $V_{\mbox{\tiny exp}}=(80\pm 4)\%$.


\section{Optimizing QKD performance}\label{Sec:OptimizingQKD}

\begin{figure}[h!]
\centering
\includegraphics[width=0.8\linewidth]{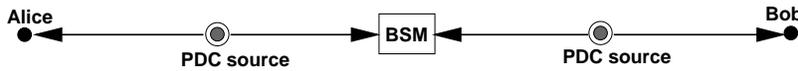}
{\caption{Illustration of ES-based QKD. 
The quantum channel between Alice and Bob is 
split into shorter segments, with two PDC sources placed $1/4$ and $3/4$ 
of the way along the channel and a joint BSM performed halfway. 
Given a successful BSM (with success probability equal to
$\frac{1}{2}\eta^2_0$), the photons arriving at
Alice and Bob are entangled despite never having interacted 
with one another, and the BBM92 protocol 
can be used to create the secret key. 
\label{Fig:AliceRelayBob}}}
\end{figure}

We numerically simulate the effect of real-world imperfections
on the two common QKD figures of merit,  quantum bit error rate 
({\small\sc QBER}) and secret key rate $R_{\rm sec}$, for an   
entanglement-based QKD experiment in which the long-distance quantum channel
(with distance $d$) between sender Alice and a receiver Bob 
is split into shorter segments with  two PDC sources placed $1/4$ and $3/4$ 
of the way along the channel and a BSM performed halfway (Fig.~\ref{Fig:AliceRelayBob}).  
Due to ES, the photons distributed between Alice and Bob are entangled, 
so the BBM92 protocol can be applied to produce the key. 
\vspace{1mm}

The {\small\sc QBER}, defined as the ratio of wrong 
bits to the total number of bits exchanged between Alice 
and Bob, is directly related to the visibility $V$ 
of four-fold coincidence measurements via the relation $\mbox{\small\sc
  QBER}=(1-V)/2$~\cite{Gisin2002}. Hence, it can be computed nonperturbatively 
using our closed-form solution~\cite{Scherer2009}. For compactness, the 
procedure for computation of $V$ is not repeated in the present paper. 

For one-way communication, which we analyze here, according to Shor and Preskill's 
security proof the secret key yield
is~\cite{ShorPreskillPRL2000}
\begin{equation}\label{Eq:Rsec}
R_{\mbox{\scriptsize sec}}=R_{\mbox{\scriptsize sift}}
\left[1-\kappa H_2(\mbox{\small\sc QBER})-H_2(\mbox{\small\sc QBER})\right].
\end{equation}
The first factor, $R_{\mbox{\scriptsize sift}}$, is the {\it sifted key} rate; 
it is the number of all coincidental detection events (per second) 
for which Alice and Bob made by chance 
compatible choices of bases in which they measured the received 
photons. Hence, the sifted key rate is only half that of the {\it raw key} rate, 
which consists of the overall number of qubits exchanged between Alice 
and Bob. The raw key rate is obtained as a product of the following 
probabilities per attempt of ES: (i) the probability that both PDCs 
emit EPPs, which is the product of their photon-pair production rates $\chi^2$, 
respectively, (ii) the probability that the generated photons arrive at the analyzers of both 
Alice and Bob as well as at the BSM device, (iii) the probabilities 
that the photons that arrive at Alice's and Bob's sites are also detected, 
and (iv) the probability that the BSM is successful, 
which is equal to $\frac{1}{2}\eta^2_0$ and thus bounded by 
its maximum value $1/2$~\cite{Calsamiglia2001}.
Hence, for the QKD scheme considered here, under our assumptions, 
\begin{equation}
R_{\mbox{\scriptsize sift}}=\frac{1}{2}\chi^2\chi^2(10^{-\alpha
  d/{40}})^4\eta^2_0(\frac{1}{2}\eta^2_0)= 
\frac{1}{4}\chi^4\eta^4_0\times 10^{-\alpha d/{10}}.
\end{equation}
The second factor of Eq.~(\ref{Eq:Rsec}) describes the effect of
privacy amplification. The two subtracted terms $\kappa H_2(\mbox{\small\sc
  QBER})$ 
and $H_2(\mbox{\small\sc QBER})$, where 
\begin{equation}
H_2(x)\equiv -x \log_2(x)-(1-x)
\log_2(1-x)\quad \mbox{for}\quad x\in[0,1]
\end{equation}
is the binary Shannon entropy function, represent 
the reduction of the key rate due to error correction and 
eavesdropping on the quantum transmission, respectively, with
$\kappa=1.22$ characterizing the efficiency of error correction algorithm 
compared to the Shannon limit~\cite{BrassardSalvail1994}.

We remark that the Shor-Preskill lower bound for the ratio 
between the number of secure key bits and the number of sifted key 
bits, as given by Eq.~(\ref{Eq:Rsec}), was derived under the assumption 
of perfect sources and detectors; i.e., it was assumed that any source 
or detector imperfections can be absorbed into eavesdropper Eve's attack.
The same bound was achieved by Koashi and Preskill in their 
QKD security proof for an arbitrary (possibly faulty) source with 
the only restriction that the source must not reveal any 
information to Eve about the basis chosen by Alice 
and Bob for their measurements~\cite{KoashiPreskillPRL2003}. 
This feature is naturally satisfied for our entanglement-based 
PDC-ES-BBM92 QKD. The Koashi-Preskill  security proof indicates 
that source defects are efficiently detected by the QKD protocol --- 
in our case the BBM92 protocol rather than BB84.  This means that 
Alice and Bob cannot be fooled into accepting a part of the secret key 
that Eve got to know by exploiting source imperfections.

Both Shor-Preskill~\cite{ShorPreskillPRL2000} 
and Koashi-Preskill~\cite{KoashiPreskillPRL2003} security proofs rely 
on the crucial assumption that Alice's and Bob's measurements are performed on qubits. 
This assumption is certainly not valid for real-world QKD.  
In our PDC-ES-BBM92 QKD scheme, polarization measurements are implemented 
by means of polarization rotators (quarter- and half-wave plates), PBSs and 
threshold detectors acting on multiphoton states in spatio-temporal optical modes. 
The corresponding detection events are theoretically described by POVMs 
over the {\em infinite}-dimensional Fock space.  
Yet, by using squashing techniques~\cite{SquashingPRL2008,SquashingPRA2010}, 
which are directly applicable to our setup, all detection events of our QKD scheme 
can indeed be reduced to a statistically equivalent two-dimensional qubit-based 
description. The existence of a squashing model permits employing the Shor-Preskill 
lower bound~(\ref{Eq:Rsec}) for our QKD scheme. It also ensures validity of   
our entanglement verification via four-fold coincidence measurements with 
(realistic) threshold detectors (see~\cite{SquashingPRA2010}).

\begin{figure}[b!]
\begin{center}
\includegraphics[width=0.995\linewidth]{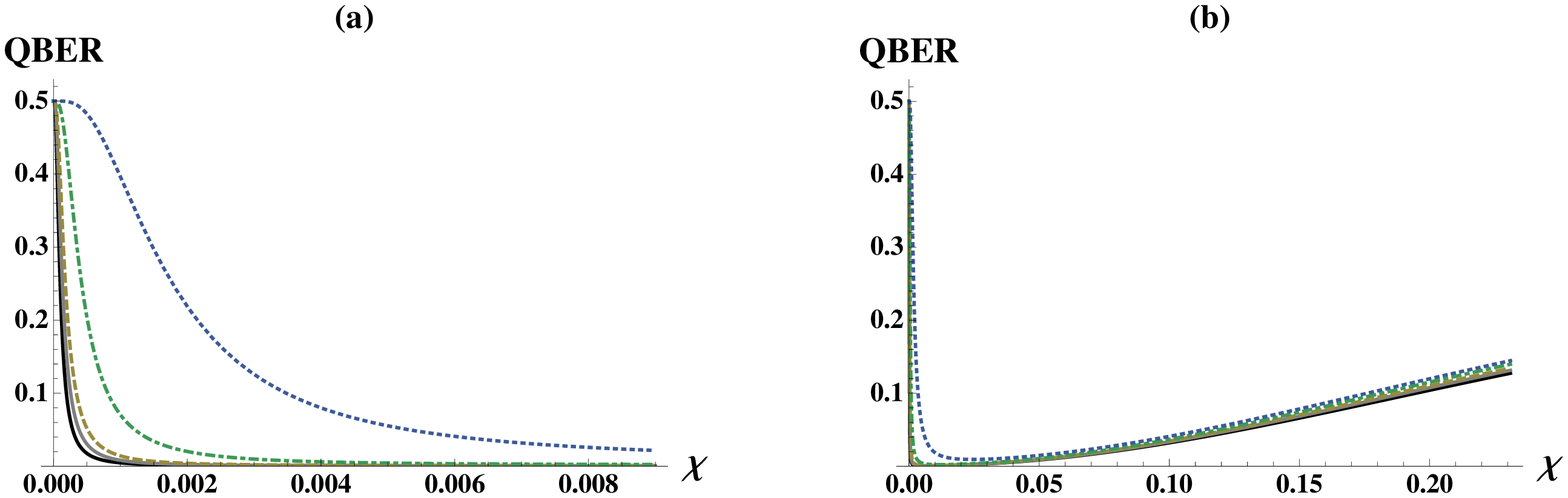} 
\includegraphics[width=0.995\linewidth]{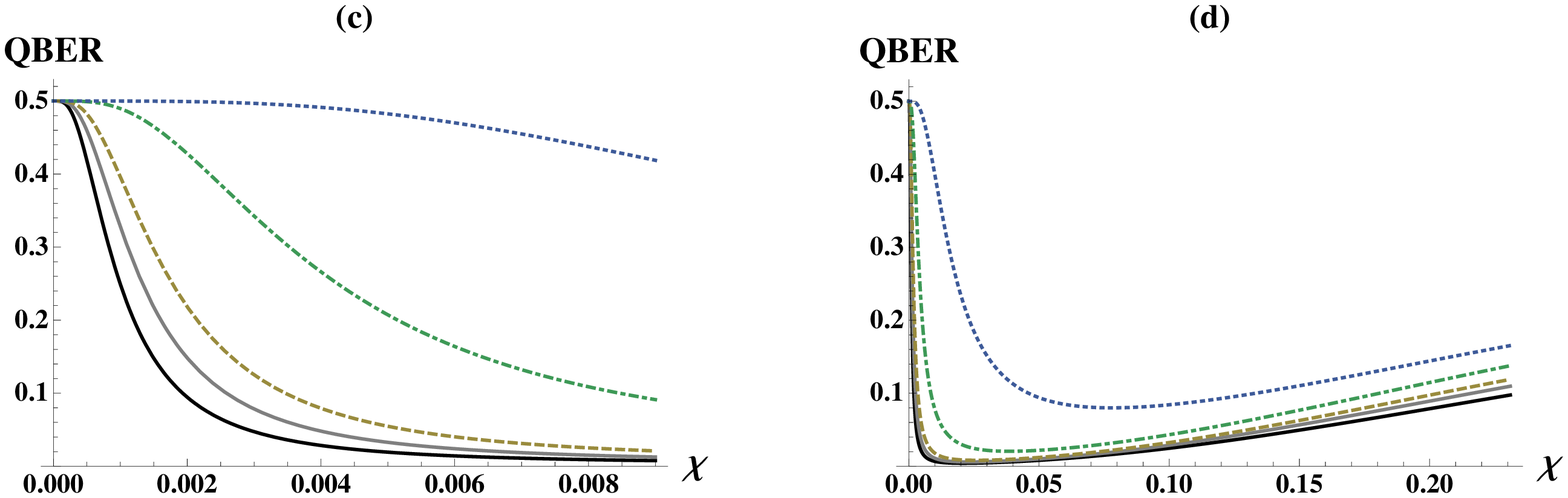} 
\end{center}
{\caption{{\small\sc QBER} as a function of $\chi$ for various
  values of the normalized distance $\alpha d$ and fixed $\eta_0$ and $\wp_{\mbox{\tiny{dc}}}$.
The function is plotted for $\alpha
  d=0,5,10,25$ and $50$, 
corresponding, respectively, to the curves of lowest- to highest-QBER
values in all diagrams, and 
$\eta_0=0.1\,$ \& $\,\wp_{\mbox{\tiny{dc}}}\approx 3\times 10^{-6}$ in figures (a)
and (b), 
or   $\eta_0=0.3$ \&  $\wp_{\mbox{\tiny{dc}}}\approx 10^{-4}$ in figures (c) and
  (d).  The values of $\eta_0$ and
  $\wp_{\mbox{\tiny{dc}}}$ are related to one another
by constraint (\ref{Eq:constraint}). 
Figures (a) and (c) display a higher resolution for very small $\chi$
  values in both cases. To have $R_{\mbox{\scriptsize sec}}>0$ 
the {\small\sc QBER} must assume values less than approx.\ $0.094$ if $\kappa=1.22$ 
($0.11$ if $\kappa=1.0$).  
\label{Fig:QBERvsChi}}}
\end{figure}

\begin{figure}[t!]
\begin{center}
\includegraphics[width=0.875\linewidth]{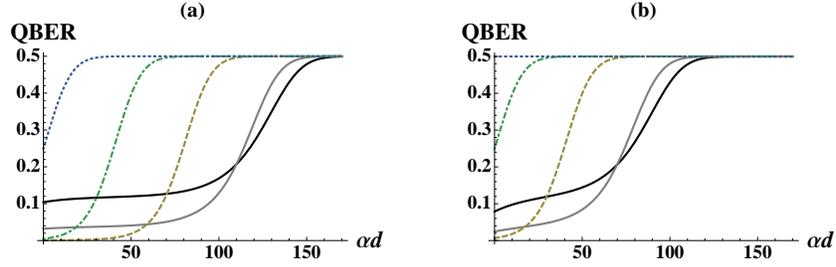} 
\end{center}
{\caption{{\small\sc QBER} vs $\alpha d$ for various
  $\chi$ values and fixed $\eta_0$ and $\wp_{\mbox{\tiny{dc}}}$.
The function is plotted for $\chi=10^{-4}$, $\,10^{-3}$, $\,10^{-2},$ $\, 0.1$, and $\,0.2$,
corresponding to the dotted, dot-dashed, dashed, gray solid  and
dark solid curves, respectively, in both diagrams, and (a)~$\eta_0=0.1\,$ \& $\,\wp_{\mbox{\tiny{dc}}}\approx 3\times 10^{-6}$ 
or (b)~$\eta_0=0.3$ \& $\wp_{\mbox{\tiny{dc}}}\approx 10^{-4}$.
The values of $\eta_0$ and $\wp_{\mbox{\tiny{dc}}}$ are related to one another
by constraint (\ref{Eq:constraint}). 
\label{Fig:QBERvsdistance}}}
\end{figure}

\begin{figure}[b!]
\centering
\includegraphics[width=0.995\linewidth]{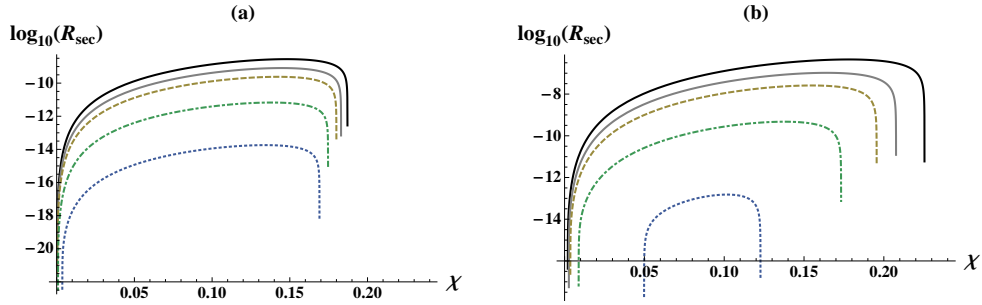} 
{\caption{$\log R_{\mbox{\scriptsize sec}}$ as a function of 
 $\chi$ for various values of the product~$\alpha d$, and fixed 
$\eta_0$ and $\wp_{\mbox{\tiny{dc}}}$. Plots are displayed 
for $\alpha d=0,5,10,25,$ and $50$, 
corresponding, respectively, to the dark solid, gray solid, dashed, dot-dashed and dotted
curves in both diagrams, 
and (a)~$\eta_0=0.1\,$ \& $\,\wp_{\mbox{\tiny{dc}}}\approx 3\times 10^{-6}$
or (b)~$\eta_0=0.3\,$  \& $\,\wp_{\mbox{\tiny{dc}}}\approx 10^{-4}$.  
Dark count probabilities are related to the values of $\eta_0$ by constraint 
(\ref{Eq:constraint}). 
The value of $R_{\mbox{\scriptsize sec}}$ is the number of secure
bits created per single pump-laser pulse.     
\label{Fig:LogSecKey}}}
\end{figure}

\begin{figure}[h!]
\centering
\begin{tabular}{cc}
\hspace*{-3mm}
\includegraphics[width=0.49\linewidth]{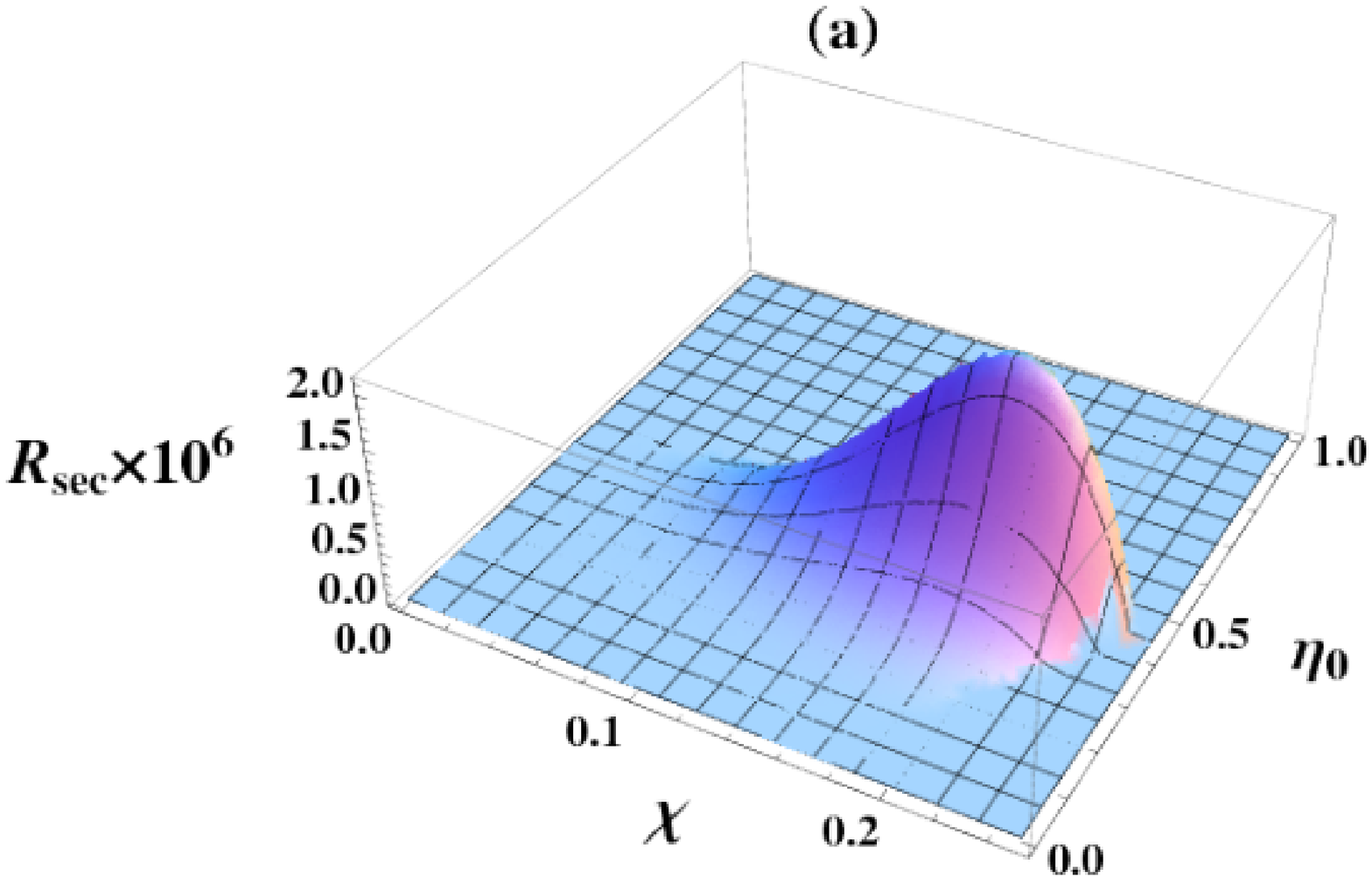} &
\hspace*{-0mm}
\includegraphics[width=0.49\linewidth]{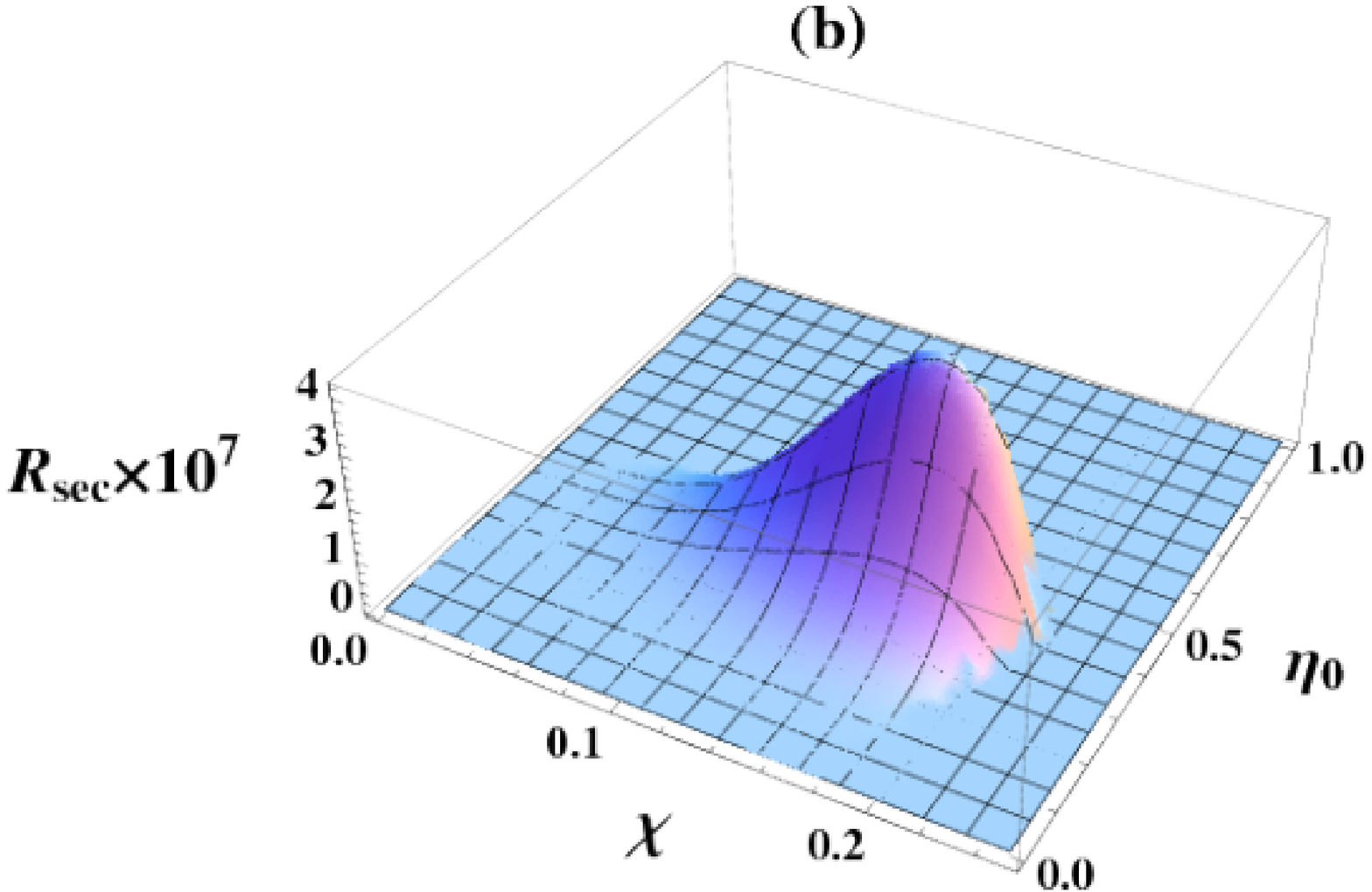} 
\end{tabular}
\vspace*{2mm}

\begin{tabular}{cc}
\includegraphics[width=0.49\linewidth]{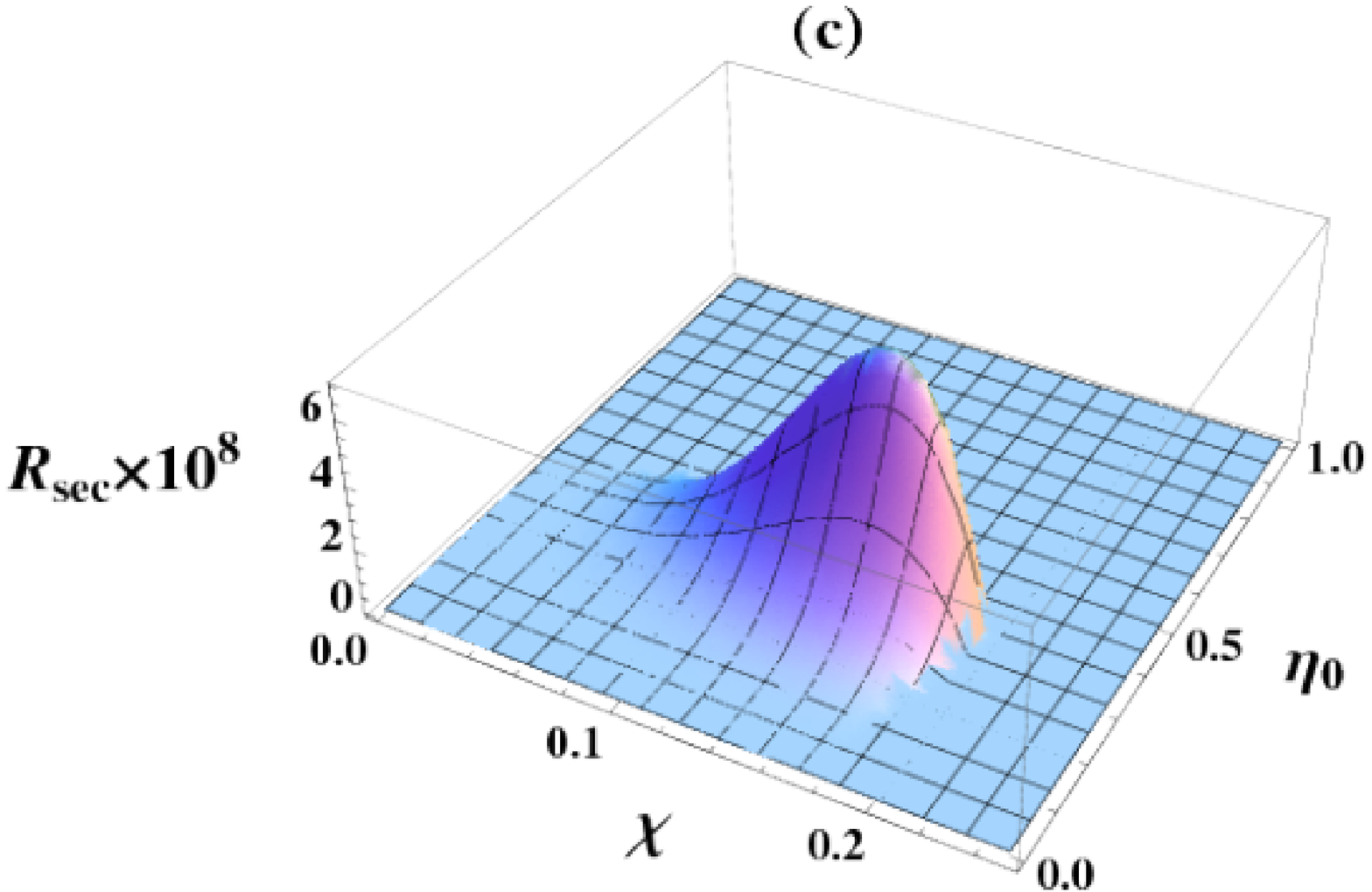} &
\hspace*{-0mm}\includegraphics[width=0.49\linewidth]{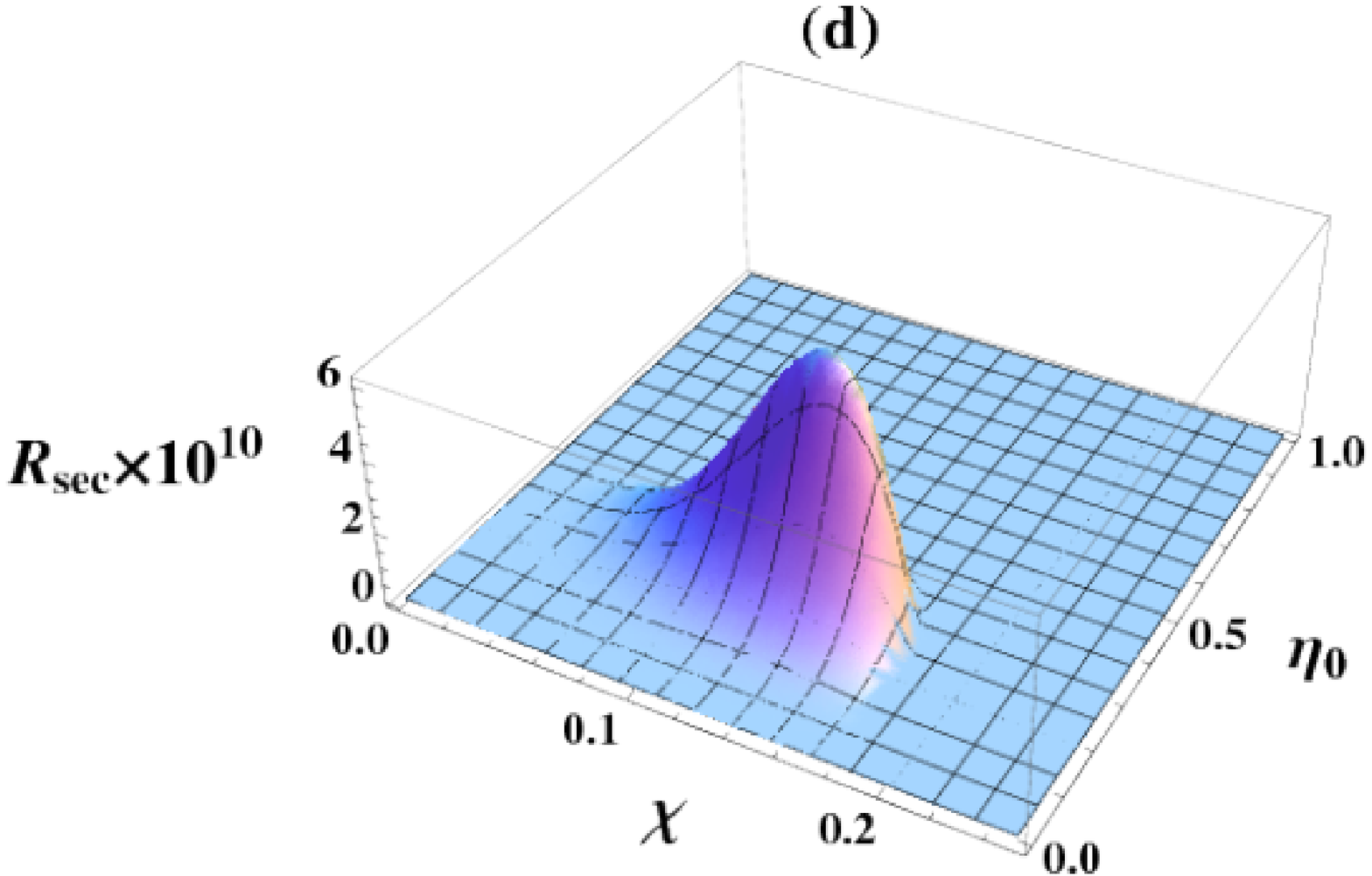} 
\end{tabular}
\vspace*{2mm}

\begin{tabular}{cc}
\includegraphics[width=0.49\linewidth]{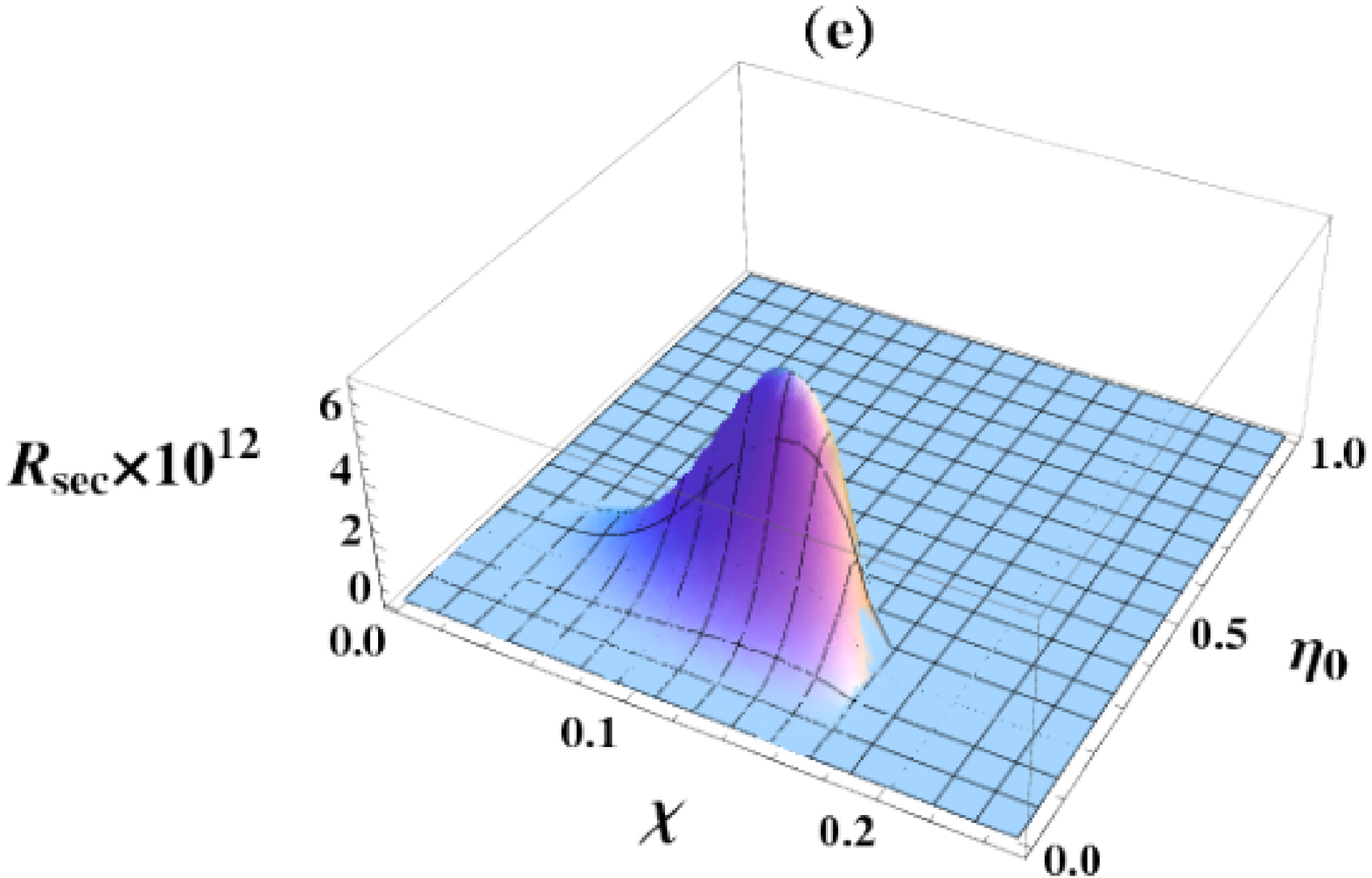} &
\hspace*{-0mm}\includegraphics[width=0.49\linewidth]{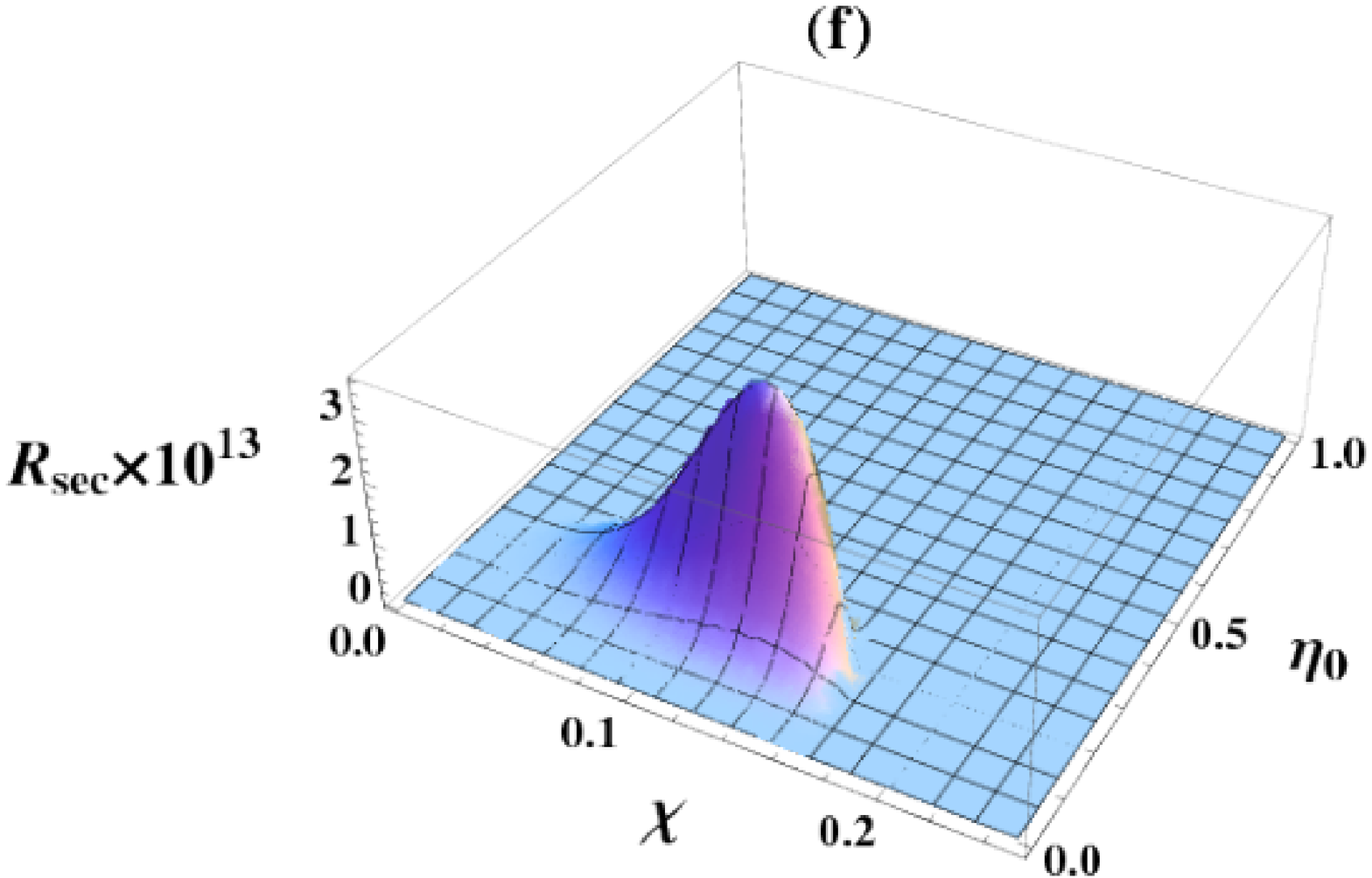} 
\end{tabular}
{\caption{Secret key rate $R_{\mbox{\scriptsize sec}}$ 
as a function 
of $\chi$ and $\eta_0$ for various exemplary values of the product~$\alpha d$. From left to right:
(a)~$\alpha d=1$, (b)~$\alpha d=5$, (c)~$\alpha d=10$,  (d)~$\alpha d=25$,
(e)~$\alpha d=40$ and (f)~$\alpha d=50$. Dark count probabilities are related 
to the values of $\eta_0$ by constraint 
(\ref{Eq:constraint}), respectively.
Here $R_{\mbox{\scriptsize sec}}$ is given in terms of the number of secure
bits created per single pump-laser pulse     
(precisely: for each attempt of ES, which requires two laser pulses, specifically with one per crystal).
\label{Fig:SecKey3D}}}
\end{figure}

\begin{figure}[h!]
\centering
\includegraphics[width=0.8\linewidth]{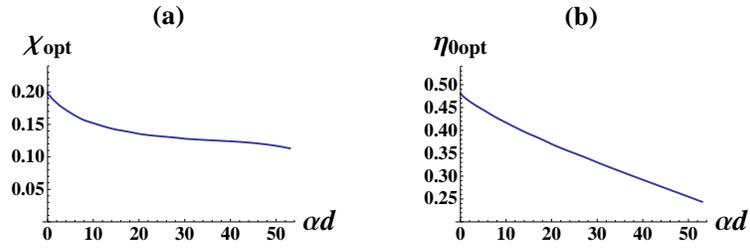}
{\caption{\small{ (a) Optimal  $\chi$ and  (b) optimal $\eta_0$ values for QKD as a
function of 
the product $\alpha d $. 
The dark count parameter $\wp_{\mbox{\tiny{dc}}}$ is related to 
$\eta_{0\mbox{\scriptsize opt}}$ by constraint~(\ref{Eq:constraint}).
\label{Fig:Optimal}}}}
\end{figure}

Our results are illustrated in
Figs.~\ref{Fig:QBERvsChi}--\ref{Fig:Optimal}.
Fig.~\ref{Fig:QBERvsChi} displays the dependence of the 
{\small\sc QBER} on the parameter $\chi$ for various 
 fixed values of the product $\alpha d$, whereas Fig.~\ref{Fig:QBERvsdistance} 
shows the {\small\sc QBER's} dependence on $\alpha d$ for various 
 fixed $\chi$ values. In both figures, $\eta_0$ and $\wp_{\mbox{\tiny{dc}}}$ 
 are fixed and interrelated by constraint~(\ref{Eq:constraint}). 
As expected, for a fixed distance, the {\small\sc QBER} 
is large for exceedingly small as well as for notably large $\chi$ values.
This dependence can be understood as follows.
In the case of excessively low photon-pair PDC production rates 
(exceedingly small $\chi$ values), most detection events   
arise due to detector dark counts, which contribute noise, 
thus implying an increase of the {\small\sc QBER}.  
As the photon-pair production rate increases, the constant detector noise
level becomes less relevant so that most detector clicks 
are due to correctly detecting single photons stemming from PDC sources,
thereby entailing a low {\small\sc QBER} value.
On the other hand, excessively  high photon-pair
production rates (large $\chi$ values) are counterproductive as they 
involve a higher probability of multipair events in the PDC process, thereby 
making the {\small\sc QBER} grow.  
As we observe in Fig.~\ref{Fig:QBERvsChi}, 
our theory predicts the value of $\chi$ that minimizes the {\small\sc
  QBER} for given channel length and loss 
coefficient. Conversely, given fixed  $\chi$, 
we know how the {\small\sc QBER} scales with distance 
$d$. We also find that, as far as {\small\sc QBER} is concerned, 
lower detector efficiency is preferable, 
given the constraint~(\ref{Eq:constraint}).   
Note that, to achieve non-vanishing secret key rates 
the {\small\sc QBER} must not exceed the value 0.094 if $\kappa=1.22$. 
(0.11 if $\kappa=1.0$).

However, optimal brightness for QKD is not given by  the value of $\chi$ that minimizes  the  
{\small\sc QBER} because two effects contribute to $R_{\mbox{\scriptsize sec}}$'s dependence
on $\chi$. The first contribution is via the sifted key rate, 
which increases proportionally with $\chi^4$. The second 
contribution is via the {\small\sc QBER} in a nontrivial way (see the second
factor in Eq.~(\ref{Eq:Rsec})).  
For the highest possible $R_{\mbox{\scriptsize sec}}$, an optimal trade-off 
between the production rate of final EPPs and the amount of entanglement after 
the ES operation has to be achieved.

For QKD the relevant quantity to be optimized is the secret key rate
$R_{\mbox{\scriptsize sec}}$, whose  
dependence on $\chi$ and $\eta_0$ is displayed in Figs.~\ref{Fig:LogSecKey} 
and \ref{Fig:SecKey3D} for various values of $\alpha d$. 
Our model reveals the  optimal $\chi$ and $\eta_0$ that maximize 
the secret key rate for given channel length. We have performed 
a constrained optimization, both with respect to $\chi$ and $\eta_0$,
assuming constraint (\ref{Eq:constraint}) between the detector efficiencies 
and dark counts for APD detectors. The result is presented in
Fig.~\ref{Fig:Optimal}.

\section{Comparison:  PDC-ES-BBM92 QKD vs Decoy-BB84 QKD}
\label{Sec:ComparisonPDC-ES-BBM92vsDecoy-BB84QKD}

Decoy-BB84 QKD, invented to combat QKD vulnerability 
due to PNS attacks, was shown to enable 
achieving much greater distances than conventional 
faint-pulse BB84 QKD without decoy states~\cite{Lo2005-1,Lo2006}. 
Furthermore, decoy-BB84 QKD realizes substantially higher 
key distribution rates than PDC-based BBM92 QKD for medium- 
and low-loss settings (i.e.~short distances), 
while the latter tolerates higher channel losses, hence 
permitting  longer communication distance than the former~\cite{Ma2007}. 
Additional range extensions are possible by means of ES 
at cost of low communication rates; this fact 
has been demonstrated for ES based on {\em ideal} EPP sources in~\cite{EdoWaks2002,Collins2005}. 
Here we compare PDC-ES-BBM92 QKD and decoy-BB84 QKD with respect to achievable 
distance. In particular, we analyze the issue whether the advantage 
of ES-based BBM92 QKD with regard to range limits is due to a better  
scaling with respect to dark counts rather than due to a higher tolerance 
of losses and thus would vanish for promising future detectors with 
negligible dark count rates. In the discussion below, 
the detector efficiency and dark counts are assumed not to be 
constrained. 

For decoy-BB84 QKD, the key generation rate is given by the formula~\cite{Lo2005-2} 
\begin{equation}
R^{\mbox{\tiny
    decoy}}_{\mbox{\scriptsize sec}}\ge \frac{1}{2}\left\{-Q_\mu f(E_\mu)
  H_2\left(E_\mu\right)+ Q_1\left[ 1- H_2\left(e_1\right)\right]\right\},
\label{Eq:Decoy1}
\end{equation}
where $\mu$ denotes the intensity  (photon number expectation
value) of signal states sent by Alice to Bob,
$Q_\mu$ is the gain of signal states, $E_\mu$ is the overall {\small\sc QBER} of signal states,
$Q_1$ is the gain of single-photon states in signal states, $e_1$ is the
error rate of single-photon states in signal states, and $f(x)$ is the error correction efficiency function. 
While $Q_\mu$ and  $E_\mu$ can be measured directly from the experiment, 
$Q_1$ and  $e_1$ have to be estimated.

Here we employ the practical vacuum \& weak-decoy state method, 
i.e., a two-decoy-state protocol with expected photon numbers $\nu_1=0$ and $\nu_2=\nu\ll
1$, which has been shown to asymptotically approach the theoretical limit 
of the most general type of decoy state protocol (with an infinite number of 
decoy states)~\cite{Lo2005-2}. See also~\cite{WangPRA2005} for an efficient 
and feasible three-decoy-state protocol (using vacuum and two decoy states).  
The weak decoy state method allows to lower-bound $Q_1$ 
and upper-bound $e_1$. Here we use the corresponding 
bounds derived in~\cite{Lo2005-2} as well as definitions of  $E_\mu$ and $Q_\mu$
provided therein. We choose the same error correction efficiency 
as for PDC-ES-BBM92 QKD in Eq.~(\ref{Eq:Rsec}), 
$f(E_\mu)=\kappa=1.22$~\cite{BrassardSalvail1994}. 
Furthermore, for a fair comparison, we assume that only dark counts and 
other background events contribute to $E_\mu$, while we neglect erroneous
detection events due to alignment and stability imperfections of the optical 
system, which have not been accounted for in our model for PDC-ES-BBM92 QKD 
either. The optimal choice of $\nu$, which depends on the transmission
distance, has also been analyzed in~\cite{Lo2005-2}; the optimal $\nu$ 
is fairly small $(\sim 0.1$) for all distances. Here we 
choose the fixed value $\nu=0.1$, which is reasonable as shown 
in~\cite{Lo2005-2}. 

\begin{figure}[t!]
\centering
\includegraphics[width=0.99\linewidth]{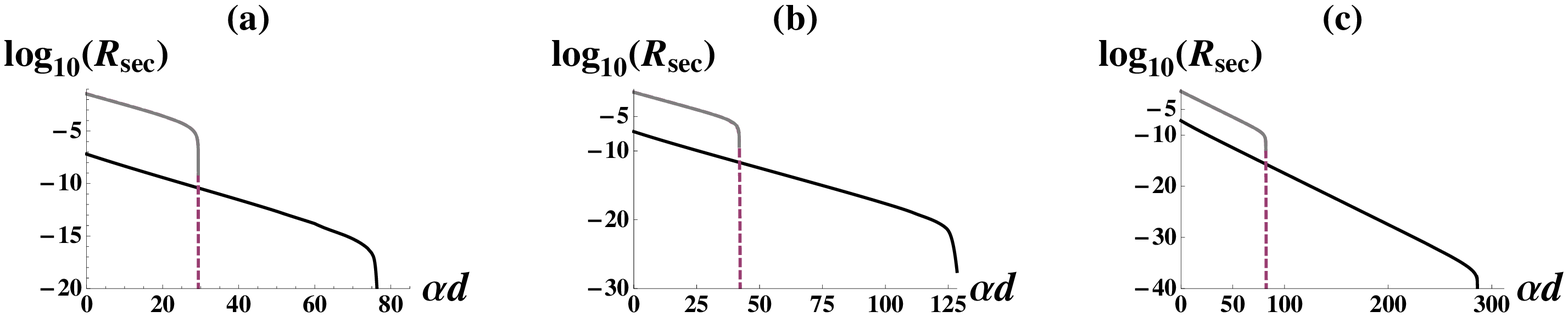} 
{\caption{Comparison between decoy-BB84 and PDC-ES-BBM92 QKD performance 
for decreasing dark count noise. All three diagrams display the 
logarithm of $R_{\mbox{\scriptsize sec}}$ vs $\alpha d$ 
for decoy-BB84 QKD (gray curve) and for PDC-ES-BBM92 QKD (dark curve), 
with optimal choice of source brightness for every value of $\alpha d$, 
respectively, and the fixed detector efficiency $\eta_0 = 0.2$. 
The detector dark count probabilities are (a)~$\wp_{\mbox{\tiny{dc}}}=1.8\times 10^{-5}$
(complying with constraint~(\ref{Eq:constraint})), (b)~$\wp_{\mbox{\tiny{dc}}}=10^{-6}$
and (c)~$\wp_{\mbox{\tiny{dc}}}=10^{-10}$  (ignoring
constraint~(\ref{Eq:constraint}) in (b) and (c)). 
\label{Fig:Comparison_for_eta20}}}
\end{figure}

We have computed the secret-key rate as a function of $\alpha d$ 
for both PDC-ES-BBM92 QKD (using Eq.~(\ref{Eq:Rsec})) and 
decoy-BB84 QKD  (using Eq.~(\ref{Eq:Decoy1}) with lower and upper bounds 
for $Q_1$ and  $e_1$ from Ref.~\cite{Lo2005-2}) for detectors with the fixed 
efficiency $\eta_0 = 0.2$ and diminishing dark count noise, see
Fig.~\ref{Fig:Comparison_for_eta20}. Optimal source brightness 
(i.e., optimal values of $\chi$ and $\mu$, respectively) has been 
chosen for each value of $\alpha d$, respectively, so as to achieve highest 
QKD performance at each distance. As demonstrated in
Fig.~\ref{Fig:Comparison_for_eta20}, decoy-BB84 QKD permits significantly 
higher key distribution rates for short distances up to the crossover point at which
$R^{\mbox{\tiny decoy}}_{\mbox{\scriptsize sec}}$ rapidly drops to
zero causing a steep slope of $\log R^{\mbox{\tiny decoy}}_{\mbox{\scriptsize
    sec}}$, while PDC-ES-BBM92 QKD enables much  longer range. 
As $\wp_{\mbox{\tiny{dc}}}$ decreases, the range of both decoy-BB84 QKD 
and PDC-ES-BBM92 QKD increases and the crossover point moves to larger
distances. As a consequence, the advantage of PDC-ES-BBM92 over decoy-BB84 QKD 
with respect to range diminishes because the communication rate of the former 
beyond the crossover point becomes gradually prohibitively low.

There is no crossover of the curves (corresponding to $\log R^{\mbox{\tiny
    decoy}}_{\mbox{\scriptsize sec}}$ and  $\log R^{\mbox{\tiny 
    ES}}_{\mbox{\scriptsize 
    sec}}$ vs $\alpha d$) for $\wp_{\mbox{\tiny{dc}}}=0$. Whereas, 
in this limit, $\log R^{\mbox{\tiny decoy}}_{\mbox{\scriptsize sec}}$
keeps decreasing linearly for all $\alpha d$, $\log R^{\mbox{\tiny 
    ES}}_{\mbox{\scriptsize 
    sec}}$ eventually drops 
exponentially. Intuitively this is obvious. Under our assumptions, in decoy-BB84 QKD 
the overall {\small\sc QBER}, $E_\mu$, and  the
error rate of single-photon states,  $e_1$, are caused only by dark count noise; 
for $\wp_{\mbox{\tiny{dc}}}=0$ both $E_\mu=0$ and $e_1=0$, implying, according
to Eq.~(\ref{Eq:Decoy1}),  $R^{\mbox{\tiny decoy}}_{\mbox{\scriptsize sec}}\ge Q_1/2$, which  
decreases exponentially without range limit. In PDC-ES-BBM92 QKD, in addition to dark counts, 
multipair events of PDC sources contribute to the overall 
quantum bit error rate, so $\wp_{\mbox{\tiny{dc}}}=0$ does not 
imply a vanishing {\small\sc QBER}. 
It is of no practical interest 
to determine the point at which the range of decoy-BB84 QKD outdistances
the range achieved in PDC-ES-BBM92 QKD, because the corresponding 
key rates become extremely small, thus useless. 

\begin{figure}[h!]
\centering
\includegraphics[width=0.9\linewidth]{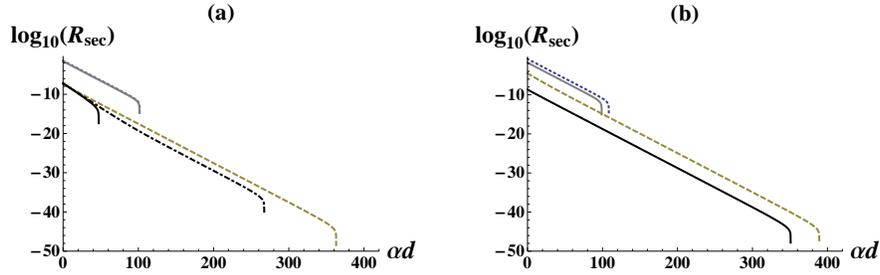} 
{\caption{(a) Effect of source brightness variation on key generation rate and 
distribution range for fixed $\eta_0=0.2$ and negligible dark counts, 
$\wp_{\mbox{\tiny{dc}}}=10^{-12}$. In decoy-BB84 QKD, 
the key rate and range are fairly stable against 
a change of $\mu$ (as already pointed out in~\cite{Lo2005-2}): 
the curves corresponding to $\mu=0.8$ (dotted)
 and $\mu=0.4$ (gray solid) coincide. In PDC-ES-BBM92 QKD, 
the key rate and achievable range depend drastically on source brightness, 
as demonstrated by the curves corresponding to  $\chi=0.174$ (dark solid), 
$\chi=0.172$ (dot-dashed) and $\chi=0.12$ (dashed), respectively. 
(b) Effect of detector efficiency variation on key generation rate and 
distribution range for fixed 
$\wp_{\mbox{\tiny{dc}}}=10^{-12}$ and fixed source brightness, $\mu=0.7$ 
and $\chi=0.120$, respectively. Predictions for two different $\eta_0$ values 
are shown. For $\eta_0=0.9$ yielding the dotted curve for decoy-BB84 
and the dashed curve for PDC-ES-BBM92 QKD, higher key rates and longer 
distances are achieved than for $\eta_0=0.1$ yielding the gray solid curve  
for decoy-BB84  and dark solid curve for PDC-ES-BBM92 QKD. The effect 
is substantially greater for PDC-ES-BBM92 than for decoy-BB84 QKD.
\label{Fig:Comparison_Dependence_chi_eta}}}
\end{figure}

As illustrated in Fig.~\ref{Fig:Comparison_Dependence_chi_eta}(a), 
the achievable range in PDC-ES-BBM92 QKD is quite sensitive to 
the choice of PDC source brightness, whereas in decoy-BB84 QKD  
the key rate and distance are fairly stable against 
a variation of $\mu$. Hence, particularly for PDC-ES-BBM92 QKD,  
it is necessary to optimize the source brightness for 
each given distance. Furthermore, for negligible 
dark counts, increasing the detector efficiency 
yields a higher key rate as well as a longer distribution 
range. The impact of detector efficiency increase on QKD performance is more 
significant for PDC-ES-BBM92 QKD than for decoy-BB84 QKD, as shown in 
Fig.~\ref{Fig:Comparison_Dependence_chi_eta}(b). 
This is easily understandable. In decoy-BB84 QKD photon detections  
take place only at Bob's site, whereas, in ES-based BBM92 QKD, 
additional detectors are employed at Alice's site as well as to 
perform a BSM.

\section{Conclusions}\label{Sec:Conclusion}

ES is a fundamental building block in
entanglement-based quantum communication schemes over 
long distances. We propose a nonperturbative theory for 
{\em practical} QKD based on ES. After identifying and 
characterizing the resources for  QKD,
we perform constrained
optimization of QKD performance with respect to PDC sources 
and detectors for any distance $d$ between sender and receiver  
and for arbitrary loss coefficients. 
For QKD schemes via a single ES operation,
the PDC brightness and the detector efficiencies  
should be tuned so that $0.12 < \chi < 0.19 $ 
and $0.25 < \eta_0 < 0.48$ depending on $\alpha d$ 
and the empirical constraint~(\ref{Eq:constraint}) between the 
detector efficiency and dark counts. 
Our theory assumes only  existing technology. 
Even though we have elaborated on PDC sources 
and APD detectors, our model can straightforwardly be 
applied to other types of realistic EPP sources and detectors. 
With respect to eavesdropping,
we assume that the  eavesdropper (Eve) 
exploits all experimental imperfections.
Our predictions provide useful upper bounds 
on the ES-based long-distance QKD performance.
Although the advantage of ES-based QKD over faint-pulse decoy-BB84 QKD, 
with respect to achievable distances, diminishes for detectors 
with negligibly small dark count rates, ES-based QKD is also important as an
enabler for quantum repeater-based QKD.

Our analysis could be further improved by  accounting for temporal-mode
overlap imperfections on a beam-splitter as well as spectral-mode mismatch. 
Moreover, we conjecture that the optimal PDC brightness  
could be shifted to higher values by employing (realistic) photon-counting  
detectors instead of threshold detectors. Our closed-form  
solution for the actual entangled quantum states prepared by practical 
ES~\cite{Scherer2009}  allows for inefficient, noisy photon-number 
discriminating detectors for the BSM. 
The intuition affirms this conjecture and is easily understood as follows. 

Let us consider the events where a coincidence detection by two threshold
detectors for modes $b$ and $c$ (in Fig.~\ref{Fig:Setup}) is interpreted 
as a projection onto a Bell state. For threshold detectors, a fraction 
of these cases originates from two (or more) photons impinging on one 
detector and (at least) one photon impinging on the other detector, 
whenever at least one PDC source has a multipair excitation. 
These harmful erroneous heralding events, 
which are not identifiable by threshold detectors, 
may result in a failure of the ES operation 
and thus increase of the {\small\sc QBER}. 
On the other hand, unit-efficiency photon-number
discriminating detectors would allow to identify 
and discard these undesired events,
which yields a lower  {\small\sc QBER} 
for equal-source brightness or, conversely, allows increasing 
brightness while keeping the {\small\sc QBER} constant.
Even imperfect photon-number discriminating detectors could reveal 
and thus enable to eliminate the erroneous heralding events to some extent. 
Hence, optimum brightness for maximal QKD 
performance could be chosen higher than in the case of threshold detectors, 
and one would achieve higher secret-key production rates due to
increasing values of the raw-key rate. This conjecture is worth examining 
in view of promising technological advancements with 
photon-number-resolving detectors~\cite{Rosenberg2005A,Rosenberg2005B,Nanowires-4}.

\vspace{1mm}
\section*{ Acknowledgements}
This project has been supported by NSERC,  
$i$CORE (now part of Alberta Innovates - Technology Futures), 
MITACS and General Dynamics Canada. BCS is a CIFAR Fellow. 
The authors thank Norbert L{\"u}tkenhaus, Eleni Diamanti, 
Romain All\'{e}aume, Ben Fortescue and Patrick Ming-yin Leung 
for helpful discussions.
\end{document}